\documentclass{Interspeech}
\usepackage{multirow}
\usepackage{booktabs}
\usepackage{tikz}              
\usepackage{array}

 \interspeechcameraready

\title{MultiAPI Spoof: A Multi-API Dataset and Local-Attention Network for Speech Anti-spoofing Detection \thanks{Corresponding Author: Ming Li}}

\author[affiliation={1}]{Xueping}{Zhang}
\author[affiliation={1}]{Zhenshan}{Zhang}
\author[affiliation={3}]{Yechen}{Wang}
\author[affiliation={3}]{Linxi}{Li}
\author[affiliation={3}]{Liwei}{Jin}
\author[affiliation={1,2}]{Ming}{Li}

\affiliation{Digital Innovation Research Center}{Duke Kunshan University}{China}
\affiliation{School of Artificial Intelligence}{The Chinese University of Hong Kong, Shenzhen}{China}
\affiliation{}{OfSpectrum, Inc.}{USA}
\email{mingli369@cuhk.edu.cn}

\keywords{Speech Anti-spoofing, Speech Deepfake Detection, MultiAPI Spoof, API Tracing, Local-Attention Network}

\usepackage{comment}

\begin{document}

\maketitle

\begin{abstract}
Existing speech anti-spoofing benchmarks rely on a narrow set of public models, creating a substantial gap from real-world scenarios in which commercial systems employ diverse, often proprietary APIs.
To address this issue, we introduce MultiAPI Spoof, a multi-API audio anti-spoofing dataset comprising about 230 hours of synthetic speech generated by 30 distinct APIs, including commercial services, open-source models, and online platforms.
Furthermore, we propose Nes2Net-LA, a local-attention enhanced variant of Nes2Net that improves local context modeling and fine-grained spoofing feature extraction. 
Based on this dataset, we also define the API tracing task, enabling fine-grained attribution of spoofed audio to its generation source.
Experiments show that Nes2Net-LA achieves state-of-the-art performance and offers superior robustness, particularly under diverse and unseen spoofing conditions.
Code \footnote{https://github.com/XuepingZhang/MultiAPI-Spoof} and dataset \footnote{https://xuepingzhang.github.io/MultiAPI-Spoof-Dataset/} have been released.
\end{abstract}

\section{Introduction}

Recently, Text-To-Speech (TTS) \cite{tts1,tts2,tts3,tts4}, Voice Conversion (VC) \cite{vc1,vc2,vc3,vc4}, and generative modeling techniques \cite{qwen2audio, gpa, gpt4,Gemini} have evolved rapidly.
In particular, end-to-end dialogue systems \cite{fun,Qwen3-omni,wavbench,Baichuan-omni}, speech continuation models \cite{Baichuan-audio, mimo-audio}, and style- or emotion-specific speech generation models \cite{SoulXPodcast,flexivoice,Step-Audio-EditX} have advanced significantly.
As a result, synthetic speech has become increasingly realistic and pervasive in everyday applications. Modern audio generation systems, especially those based on diffusion and large-scale generative models \cite{diffu, Qwen3-omni, Kimi-audio}, can now produce speech that closely mimics human prosody, timbre, and emotion. However, they have introduced serious security risks and can be easily misused for impersonation or misinformation.

Recent audio anti-spoofing approaches are typically based on a pre-trained model \cite{xlsr, hubert, wavlm} that extracts high-level acoustic features. These features are then fed into a back-end classifier \cite{res2net, aasist, mamba,nes2net} to distinguish bona fide from spoofed audio. Although recent studies in audio anti-spoofing have achieved notable progress, existing datasets \cite{add,sstc, asv19, asv21, asv5,codecfake} are typically constructed from a limited number of public TTS or VC models, providing an incomplete view of today’s real-world spoofing landscape. In practice, most industrial platforms adopt proprietary or closed-source APIs, making it difficult to access their model architectures, data pipelines, or synthesis mechanisms. Therefore, it remains unclear how well models trained on existing open-source benchmarking datasets will perform on real-world API data. Moreover, the rapid emergence of new generative paradigms results in a substantial domain gap between research benchmarks and real-world spoofing attacks.

To address these limitations, we introduce MultiAPI Spoof, a new multi-API speech anti-spoofing dataset designed for both anti-spoofing detection and API-level source tracing. Unlike prior datasets that focus on a few synthesis systems, MultiAPI Spoof comprises audio generated from 30 distinct APIs, including commercial TTS services, open-source speech models, and TTS websites. The dataset covers approximately 230 hours of spoofed speech. Based on the MultiAPI Spoof dataset, our contributions are as follows:
\begin{enumerate}
    \item We show that there is a gap between previous research benchmarks and real-world spoofing scenarios; adding our API dataset in the training can also enhance the performance on current benchmarks.
    \item Furthermore, we propose a new anti-spoofing detection method, namely Nes2Net-LA, built upon Nes2Net \cite{nes2net}. By integrating local attention modules between Nested blocks, Nes2Net-LA enhances local context modeling and fine-grained spoofing feature extraction, thereby improving robustness and discriminative capability. The Nes2Net-LA achieves state-of-the-art (SOTA) performance across multiple anti-spoofing benchmarks.
    \item Finally, we introduce the API tracing task, which aims to identify the generation API of spoofed audio and establishes a benchmark for fine-grained source attribution.
\end{enumerate}

\section{MultiAPI Spoof Dataset}

MultiAPI Spoof is a new multi-API audio anti-spoofing dataset designed to bridge the gap between research benchmarks and real-world synthetic speech. It contains approximately 230 hours of spoofed audio and an equal amount of bona fide speech from CommonVoice, maintaining a 1:1 balance between the two. All recordings are in English. The dataset provides a diverse set of spoofing conditions for both anti-spoofing detection and API-level source tracing.

\subsection{Spoofed Audio Data Sources}

The spoofed audio in MultiAPI Spoof is generated through 30 distinct APIs, reflecting a broad spectrum of synthesis techniques and real-world deployment scenarios:
\begin{enumerate}
\item Commercial TTS APIs: Speech synthesized by proprietary text-to-speech services widely used in industry.
\item Open-Source Models: Speech generated using publicly available neural TTS or voice conversion systems.
\item TTS Websites: Audio collected from online platforms providing web-based speech synthesis interfaces.
\end{enumerate}
Each API corresponds to one labeled group (A0–A29), forming a comprehensive representation of modern TTS and generative pipelines.

\subsection{Dataset Split}

The MultiAPI Spoof dataset is partitioned by the APIs. APIs A0--A20 are used to construct the training, development, and evaluation subsets with a 70/10/20 \% split, ensuring sufficient variation within seen sources. APIs A21--A23 are reserved entirely for development, while APIs A24--A29 are held out exclusively for evaluation. This design enables two evaluation conditions:
\begin{enumerate}
\item Seen evaluation, where systems are tested on spoofed samples generated from APIs that also appear in training (A0--A20).
\item Unseen evaluation, where systems are evaluated on spoofed samples from completely unseen APIs (A21--A29), allowing assessment of cross-source generalization.
\end{enumerate}

\section{Local Attention Enhanced Anti-spoofing Network}

\begin{figure}[t]
\centerline{\includegraphics[width=0.5\textwidth]{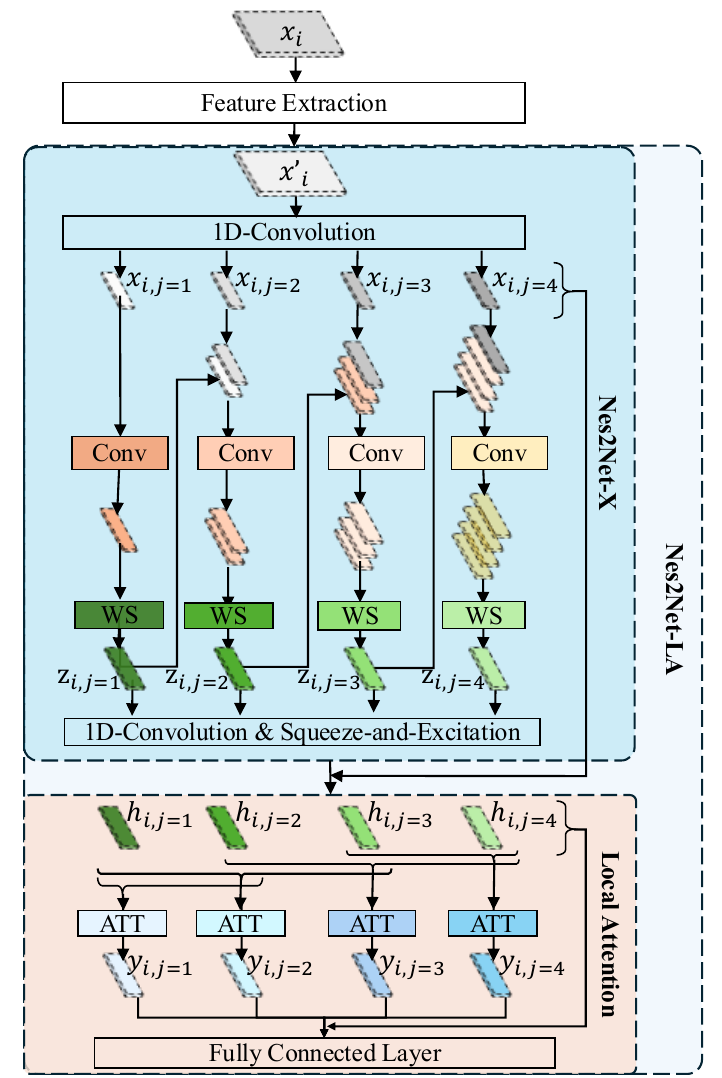}}
\caption{Overall architecture of proposed  Nes2Net-LA frameworks.
The model first extracts high-dimensional representations from the input audio and then processes them using nested multi-scale feature fusion. Nes2Net-LA further enhances cross-block interactions through a sliding-window local attention mechanism. `WS' represent Weighted Summation, and `ATT' represent scaled dot-product self-attention.}
\label{main}
\end{figure}

\subsection{Prior Knowledge: Nested Res2Net (Nes2Net-X)}

The Nes2Net-X architecture \cite{nes2net} is a multi-scale feature extractor for high-dimensional speech representations. An audio segment $x_i$ is encoded into $x_i' \in \mathbb{R}^{C\times T'}$ and split into channel-wise subsets $x_{i,1},\dots,x_{i,J}$. Each subset is processed hierarchically: the first passes through Convolution(Conv) and Weighted Summation (WS), while subsequent subsets are fused with the previous output before convolution. All outputs are refined with a convolution and Squeeze-and-Excitation (SE) module with residual connections to obtain the anti-spoofing feature representations $h_{i,j} \in R^{(C/J) \times T'}$, as shown in \eqref{x1}.
\begin{equation}
\begin{aligned}
z_{i,j} &= \mathrm{WS}\Big(\mathrm{Conv}\big(x_{i,j} + \mathbb{I}_{\{j>1\}}\, z_{i,j-1}\big)\Big), \\
h_{i,j} &= x_{i,j} + \mathrm{SE}\Big(\mathrm{Conv}(z_{i,j})\Big)
\end{aligned}
\label{x1}
\end{equation}

\begin{table*}[t]
\caption{Comparison of anti-spoofing performance without and with MultiAPI Spoof training set in training. 
Each cell in the table follows the format EER$\downarrow$ / minDCF$\downarrow$ / actDCF$\downarrow$.
The `wo MultiAPI Spoof' setting trains models only on TIMIT, ODSS, FoR, AI4T, ASV5, and MLAAD, without any MultiAPI Spoof training set.
The `with MultiAPI Spoof' setting trains on the same data, plus the MultiAPI Spoof training set.}
\label{tab:multiapi_contribution}
\setlength{\tabcolsep}{2pt}

\centering
{\fontsize{8.5pt}{10pt}\selectfont
\begin{tabular}{c|c|c|ccc|c}

\hline
\multirow{2}{*}{\textbf{Dataset}}                                                  & \multirow{2}{*}{\textbf{Model}}                                           & \multirow{2}{*}{\textbf{ITW}} & \multicolumn{3}{c|}{\textbf{MultiAPI Spoof}}                                                                  & \multirow{2}{*}{\textbf{AI4T}} \\ \cline{4-6}
                                                                                   &                                                                           &                               & \multicolumn{1}{c|}{\textbf{Seen}}         & \multicolumn{1}{c|}{\textbf{Unseen}}      & \textbf{Overall}     &                                \\ \hline
\multirow{3}{*}{\begin{tabular}[c]{@{}c@{}}without \\ MultiAPI \\ Spoof\end{tabular}}   & XLSR+AASIST \cite{xlsr+aasist}                                                           & 2.02 / 0.026 / 0.029          & \multicolumn{1}{c|}{-}                     & \multicolumn{1}{c|}{-}                    & 7.30 / 0.098 / 0.106 & 12.96 / 0.132 / 0.190          \\
                                                                                   & XLSR+Nes2Net \cite{nes2net}                                                             & 1.73 / 0.023 / 0.025          & \multicolumn{1}{c|}{-}                     & \multicolumn{1}{c|}{-}                    & 7.08 / 0.098 / 0.103 & ~~7.77 / 0.093 / 0.110           \\
                                                                                   & \textbf{\begin{tabular}[c]{@{}c@{}}XLSR+Nes2Net-LA\\ (Ours)\end{tabular}} & 1.70 / 0.023 /0.020            & \multicolumn{1}{c|}{-}                     & \multicolumn{1}{c|}{-}                    & 6.11 / 0.085 / 0.089 & 7.76 / 0.090 / 0.099                               \\ \hline
\multirow{3}{*}{\begin{tabular}[c]{@{}c@{}}with \\ MultiAPI \\ Spoof\end{tabular}} & XLSR+AASIST \cite{xlsr+aasist}                                                            & 2.09 / 0.028 / 0.030          & \multicolumn{1}{c|}{\textbf{0.48} / 0.007 / 0.0070} & \multicolumn{1}{c|}{0.83 / 0.010 / 0.012} & 0.70 / 0.009 / 0.010 & ~~6.26 / 0.079 / 0.092           \\
                                                                                   & XLSR+Nes2Net \cite{nes2net}                                                             & 1.69 / 0.024 / 0.024          & \multicolumn{1}{c|}{0.55 / 0.007 / 0.008}  & \multicolumn{1}{c|}{0.80 / 0.011 / 0.012} & 0.69 / 0.010 / 0.010 & ~~\textbf{5.64} / 0.052 / 0.079           \\
                                                                                   & \textbf{\begin{tabular}[c]{@{}c@{}}XLSR+Nes2Net-LA\\ (Ours)\end{tabular}} & \textbf{1.42} / 0.020 / 0.021          & \multicolumn{1}{c|}{\textbf{0.48} / 0.007 / 0.007}  & \multicolumn{1}{c|}{\textbf{0.62} / 0.009 / 0.009} & \textbf{0.56} / 0.008 / 0.008 & ~~\textbf{5.64} / 0.051 / 0.077           \\ \hline
\end{tabular}
}
\end{table*}
\subsection{Nes2Net with Local Attention (Nes2Net-LA)}
While Nes2Net-X \cite{nes2net} effectively captures multi-scale structures, its refinement remains strictly hierarchical: each nested block only interacts with its immediate predecessor. This constrains long-range communication across blocks, which is increasingly important for high-dimensional speech representations. Hence, we propose to add Local Attention to the Nes2Net model (Nes2Net-LA).

As shown in Figure \ref{main}, for each block, we define a local sliding-window neighborhood 
$\mathcal{N}(i,j) = \{h_{i,k} \mid k \in [j-K, j+K]\}$, where $K$ is the window radius.
A local scaled dot-product self-attention (ATT) \cite{att} operator is then applied to get the local feature representation $y_{i,j}  \in \mathbb{R}^{(C/J) \times T'}$, as shown in \eqref{la}.
\begin{eqnarray}
\begin{array}{c}
y_{i,j} = \mathrm{ATT}\big(h_{i,j},\ \mathcal{N}(i,j)\big),
\end{array}
\label{la}
\end{eqnarray}
Finally, a residual connection aggregates the original representation $h_{i,j}$ and enhanced local feature representation $y_{i,j}$. All block outputs are concatenated and fed to a fully connected (FC) layer to produce the final anti-spoofing score.

Unlike global attention, which is too expensive for extended sequences of nested blocks, the proposed local attention considers only a small sliding window of neighboring blocks (e.g., 3). Within this window, each block can gather useful information from its nearby blocks and combine it with its own features. This approach makes the features more consistent and robust, improving the model's overall performance in anti-spoofing tasks.

\section{Anti-Spoofing API Tracing Task}

The anti-spoofing API tracing task aims to identify which API generated a given spoofed audio sample. Unlike conventional anti-spoofing, which only distinguishes bona fide and spoofed speech, API tracing provides fine-grained attribution. 
APIs are divided into seen and unseen sets. The seen set, consisting of 21 APIs (A0–A20), appears in training, while the unseen set is reserved for evaluation to test generalization.

Our baseline model uses hidden representations from the XLSR-300M \cite{xlsr} encoder, followed by an attention pooling layer to aggregate embeddings from each encoder hidden layer, and ends with a Squeeze-and-Excitation (SE) layer to get the final results. 
During training, only the 21 seen APIs are used. At inference, samples whose maximum predicted probability falls below a threshold are classified as the unseen class, effectively turning the task into a 22-class classification problem. 

\section{Experiments}
\subsection{Experimental Setup}
\textbf{Dataset}
The anti-spoofing experiments are conducted on a collection of six public datasets: TIMIT \cite{TIMIT}, ODSS \cite{ODSS}, FoR \cite{FoR}, AI4T \cite{AI4T}, ASV5 \cite{asv5}, and MLAAD \cite{MLAAD}. These corpora cover a wide variety of spoofing sources, including real-world collected data, text-to-speech (TTS), and voice conversion (VC). We consider two training configurations. In the first setting, the six datasets are merged into a single training set, and evaluation is performed across three target domains: the full ITW dataset \cite{ITW}, MultiAPI Spoof test set, and AI4T test set. In the second setting, the MultiAPI Spoof training set is additionally included in the training pool, and the models are re-evaluated on the same three domains.

For the API tracing task, both training and testing are performed on MultiAPI Spoof, and we report separate results for seen and unseen API categories to reflect generalization across API sources.

\textbf{Processing}
All systems operate on normalized raw waveforms. Each audio sample is converted into a 4-second segment: signals shorter than 4 seconds are repeated until reaching the 4-second length, and longer signals are truncated. Unlike many existing audio anti-spoofing systems, we do not apply data augmentation in any of our experiments to ensure a clean, controlled comparison across models.

\textbf{Training}
The anti-spoofing models evaluated in this work include XLSR+AASIST \cite{xlsr+aasist}, XLSR+Nes2Net-X \cite{nes2net}, and XLSR+Nes2Net-LA. All of those models have the same feature extroctor XLSR-300M \cite{xlsr}. For both Nes2Net-X and Nes2Net-LA, the number of channel splits is fixed at $J=8$, and the local attention module uses a window size of $K=1$. During training, XLSR+AASIST is optimized using Adam \cite{adam} with an initial learning rate of $1\times10^{-6}$, weight decay of $1\times10^{-4}$, and cross-entropy loss \cite{ce}. The XLSR+Nes2Net-X and XLSR+Nes2Net-LA systems use Adam with an initial learning rate of $5\times10^{-6}$ and weight decay of $1\times10^{-4}$, and cross-entropy loss.

For the API tracing experiments, the model is trained using Adam with a learning rate of $1\times10^{-5}$, weight decay of $1\times10^{-4}$, and cross-entropy loss.

\textbf{Metrics}
For the anti-spoofing task, performance is evaluated using Equal Error Rate (EER$\downarrow$), minimum Decision Cost Function (minDCF$\downarrow$), and actual Decision Cost Function (actDCF$\downarrow$) \cite{DCF}. For the API tracing task, we measure classification performance using precision, recall, and F1 \cite{f1}. Specifically, the F1 for seen APIs is computed as the macro-average of the F1 scores over the 21 seen API classes. For unseen APIs, F1 is computed on the single unseen class. The overall performance is reported as the macro-average across all classes, including the unseen class.

\subsection{Experimental Results and Analysis}

\subsubsection{Anti-Spoofing on MultiAPI Spoof}

To assess the value of MultiAPI Spoof for training and evaluation, we design two comparison experiments. The results are shown in Table~\ref {tab:multiapi_contribution}.

In the first comparison experiment, models are trained on six commonly used anti-spoofing datasets (TIMIT \cite{TIMIT}, ODSS \cite{ODSS}, FoR \cite{FoR}, AI4T \cite{AI4T}, ASV5 \cite{asv5}, MLAAD \cite{MLAAD}) without including MultiAPI Spoof training set. Two systems, XLSR+AASIST \cite{xlsr+aasist} and XLSR+Nes2Net-X \cite{nes2net}, are evaluated on ITW \cite{ITW}, MultiAPI Spoof test set, and AI4T \cite{AI4T}, respectively. 
Both models present relatively high EERs on the MultiAPI Spoof evaluation set, indicating a domain shift that existing datasets fail to cover.
In the second comparison experiment, MultiAPI Spoof training set is incorporated into the training pool. Across all evaluation sets, especially on the MultiAPI Spoof test set itself, both XLSR+AASIST and XLSR+Nes2Net-X show substantial reductions in EER, minDCF, and actDCF. For instance, XLSR+AASIST decreases the EER from 7.30\% to 0.70\% on MultiAPI Spoof test set, and similarly, XLSR+Nes2Net decreases the EER from 7.08\% to 0.69\%.

Moreover, the benefits are not limited to MultiAPI Spoof. On the ITW dataset, XLSR+Nes2Net decreases from 1.73\% to 1.69\% in EER, and on AI4T, it decreases from 7.77\% to 5.64\%. Notably, within MultiAPI Spoof test set itself, the gains are observed not only on the seen sources but also on the unseen subset, indicating that the additional spoofing conditions contribute to more robust feature learning rather than overfitting to specific APIs. These consistent improvements across multiple evaluation sets suggest that adding MultiAPI Spoof in the training effectively enhances cross-domain robustness and provides better generalization to unseen data.

\begin{figure}[t]
\centerline{\includegraphics[width=0.45\textwidth]{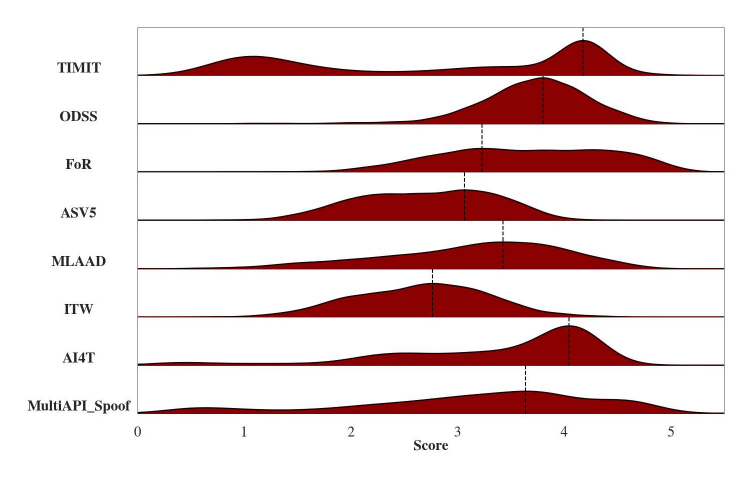}}
\caption{Scoreq \cite{Scoreq} distribution comparison across datasets.
The dashed vertical line in each curve marks the peak density value.}
\label{mos}
\end{figure}

To better understand this effect, we visualize the Scoreq\cite{Scoreq} distributions of those datasets, as shown in Figure \ref{mos}. MultiAPI Spoof exhibits a significantly broader quality distribution, spanning both low- and high-quality regions. Such diversity helps the model generalize better by preventing overfitting to narrow acoustic conditions, thereby improving detection performance in more realistic, heterogeneous environments.

\begin{table}[t]
\caption{Comparison of our proposed XLSR+Nes2Net-LA system with recent state-of-the-art anti-spoofing models.
“SP” denotes Sample Pruning \cite{AI4T}; “RB” denotes RawBoost augmentation \cite{Rawboost}; “C” denotes codec augmentation.
“Data Collection 1” consists of ASVspoof 2019 \cite{asv19}, FoR\cite{FoR}, ASVspoof 2021 DF\cite{asv21}, TIMIT\cite{TIMIT}, ODSS\cite{ODSS}, MLAAD\cite{MLAAD}, and ASV5\cite{asv5} training sets.
On top of Data Collection 1, Data Collection 2 replaces ASVspoof 2019 and ASVspoof 2021 DF with high-quality AI4T \cite{AI4T} and the proposed MultiAPI Spoof training set.
}
\label{sota}
\setlength{\tabcolsep}{3pt}
\renewcommand{\arraystretch}{1.1}
{\fontsize{8.5pt}{10pt}\selectfont
\begin{tabular}{@{}ccccc@{}}
\hline
\textbf{Model}                                                    & \textbf{Training data} & \textbf{Aug.}                                        & \textbf{ITW}  & \textbf{AI4T} \\ \hline
XLSR+SLS \cite{sls}                                                         & ASVspoof 2019 LA       & RB                                                   & 7.46          & N/A           \\ \hline
XLSR+Mamba \cite{mamba}                                                     & ASVspoof 2019 LA       & RB                                                   & 6.71          & N/A           \\ \hline
XLSR+AASIST  \cite{xlsr+aasist}                                                     & ASVspoof 2019 LA       & RB                                                   & 10.46         & N/A           \\
XLSR+AASIST \cite{xlsr+aasist}                                         & Data Collection 2      & N/A                                                  & 2.09          & 6.26          \\ \hline
XLSR+LRC  \cite{AI4T}                                                       & ASVspoof 2019          & N/A                                                  & 3.4           & 27.4          \\
XLSR+LRC   \cite{AI4T}                                              & Data Collection 1      & SP                                                   & 1.70          & 12.4          \\   \multicolumn{5}{@{}c@{}}{%
  \vspace{-0.8ex}
}\\[-0.8ex]
XLSR+LRC  \cite{AI4T}                                              & Data Collection 1      & \begin{tabular}[c]{@{}c@{}}SP \&\\ RB+C\end{tabular} & 1.90          & 10.2          \\ \hline
XLSR+Nes2Net \cite{nes2net}                                                     & ASVspoof 2019          & RB                                                   & 5.52          & N/A           \\
XLSR+Nes2Net  \cite{nes2net}                                         & Data Collection 2      & N/A                                                  & 1.69          & \textbf{5.64} \\ \hline
\begin{tabular}[c]{@{}c@{}}XLSR+\\ Nes2Net-LA (Ours)\end{tabular} & Data Collection 2      & N/A                                                  & \textbf{1.42} & \textbf{5.64} \\ \hline
\end{tabular}
}
\end{table}

\subsubsection{Effectiveness of Local Attention (Nes2Net-LA)}
As shown in Table~\ref{tab:multiapi_contribution} and Table~\ref{sota}, Nes2Net-LA trained on the data collection pool outperforms recent state-of-the-art models across all evaluation benchmarks, even without any data augmentation or pruning. The most substantial improvements are observed on the unseen split of the MultiAPI Spoof test set. These results demonstrate that the proposed local attention mechanism produces more discriminative and robust anti-spoofing representations.

\subsubsection{API Tracing on MultiAPI Spoof}

Table~\ref{tab:api} presents the API tracing results on MultiAPI Spoof. The model is trained on the MultiAPI Spoof training set and tested on the dev and eval sets. We report results separately for seen and unseen API types. Overall performance is high across seen APIs, with both dev and eval achieving substantial precision, recall, and F1 scores. However, high precision but low recall for the unseen class shows that predictions are accurate, but many unseen-class instances are not correctly identified and are falsely rejected as unseen cases. This indicates that current methods for this task, particularly in handling unseen APIs, still require further investigation.

\begin{figure}[t]
\centerline{\includegraphics[width=0.48\textwidth]{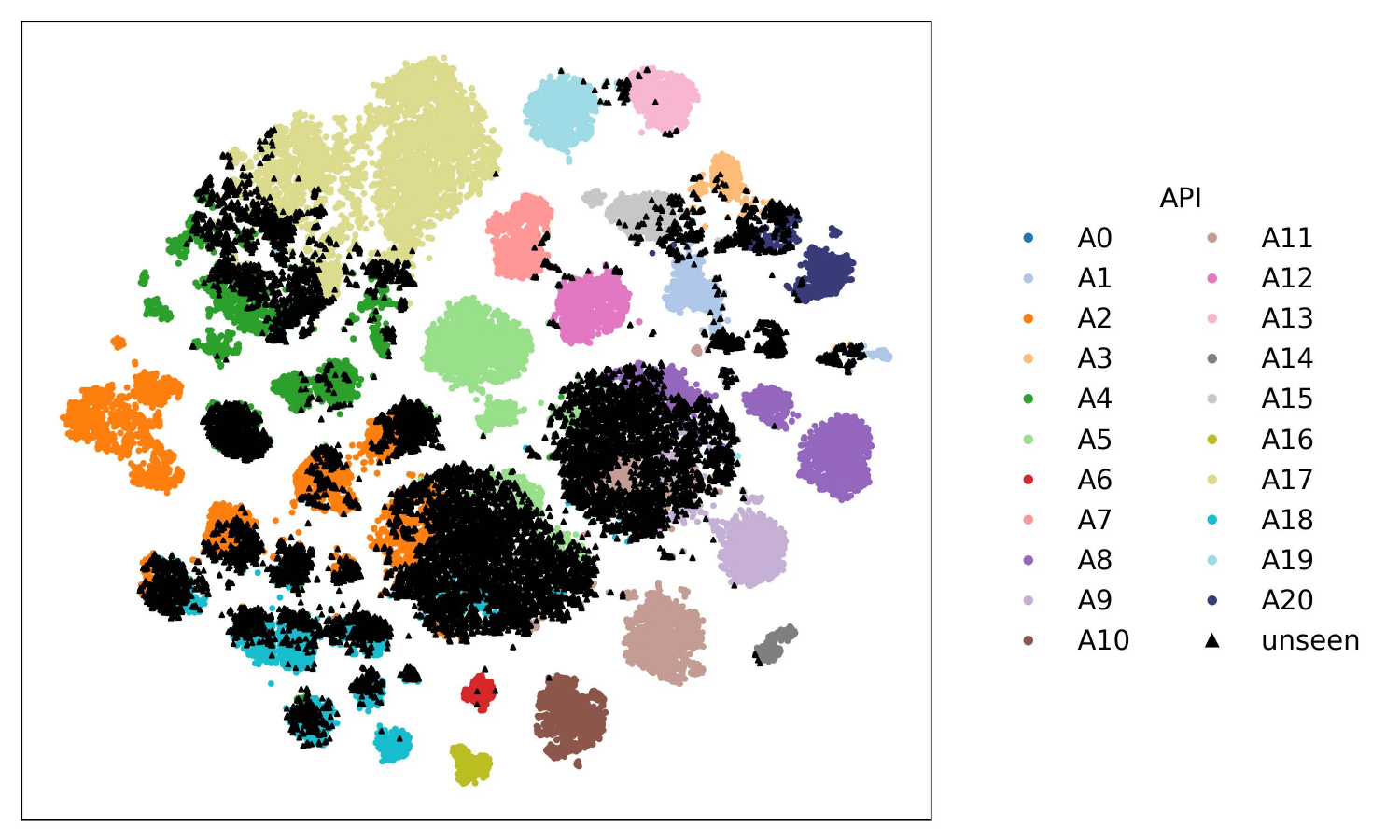}}
\caption{t-SNE \cite{tsne} visualization of XLSR-extracted embeddings for the MultiAPI Spoof eval set.
Unseen APIs are A24-A29}
\label{api}
\end{figure}

As shown in Figure~\ref{api}, t-SNE \cite{tsne} visualizations further reveal that embeddings of unseen APIs do not form separable clusters; instead, they are mixed with multiple seen categories. This suggests that the model primarily learns API-specific acoustic cues and struggles to generalize to unseen APIs whose acoustic or behavioral signatures differ significantly from the training distribution. These findings highlight the challenge of zero-shot API tracing and suggest that future models require stronger invariant representation learning. 

\begin{table}[t]
\caption{API Tracing Performance on the MultiAPI Spoof Dataset. Seen APIs correspond to A0–A20, while dev unseen APIs are A21–A23, and eval unseen APIs are A24–A29.}
\begin{tabular}{@{}c@{}|@{}c@{}cc|@{}c@{}cc@{}}
\hline
\multirow{2}{*}{} & \multicolumn{3}{c|}{\textbf{dev}}            & \multicolumn{3}{c}{\textbf{eval}}            \\ \cline{2-7} 
                  & \textbf{precision$\uparrow$} & \textbf{recall$\uparrow$} & \textbf{F1$\uparrow$} & \textbf{precision$\uparrow$} & \textbf{recall$\uparrow$} & \textbf{F1$\uparrow$} \\ \hline
\textbf{seen}     & 0.950        & 0.924           & 0.937       & 0.950        & 0.923           & 0.936       \\
\textbf{unseen}   &\textbf{ 0.959}        & 0.467           & 0.628       & \textbf{0.972}        & 0.520           & 0.678       \\
\textbf{overall}  & 0.778        & 0.910           & 0.785       & 0.770        & 0.917           & 0.782       \\ \hline
\end{tabular}
\label{tab:api}
\end{table}

\section{Conclusion}
In this paper, we present MultiAPI Spoof, a multi-API speech anti-spoofing dataset, and further introduce the API tracing task for fine-grained source attribution.
Experiments show that incorporating MultiAPI Spoof into training significantly improves cross-domain robustness.
We also propose a local-attention enhanced anti-spoofing network, namely Nes2Net-LA. It outperforms Nes2Net-X, achieving state-of-the-art performance, demonstrating its effectiveness in improving robustness and discriminative capability.

\clearpage

\section{Generative AI Use Disclosure }
Large LanguageModels (LLMs) were used solely for manuscript polishing (e.g., rephrasing and grammar checks) to improve clarity and readability. The LLMs were not used for ideation, methodology, experimental design, data analysis, or result interpretation. All scientific content was produced and verified by the authors.

\section{Acknowledgments}
Many thanks for the computational resource provided by the Advanced Computing East China Sub-Center.

\bibliographystyle{IEEEtran}
\bibliography{mybib}

\end{document}